\documentclass[twocolumn,prl,amsmath,amssymb,showpacs]{revtex4}
\usepackage{graphicx}
\usepackage{dcolumn}
\usepackage{bm}

\begin{document}

\title{Buckling Thin Disks and Ribbons with Non-Euclidean Metrics}
\author{Christian D. Santangelo}
\affiliation{Department of Physics, University of Massachusetts, Amherst, MA  01003, USA}
\email{csantang@physics.umass.edu}
\date{\today}

\begin{abstract}
I consider the problem of a thin membrane on which a metric has been prescribed, for example by lithographically controlling the local swelling properties of a polymer thin film. While any amount of swelling can be accommodated \textit{locally}, geometry prohibits the existence of a global strain-free configuration. To study this geometrical frustration, I introduce a perturbative approach. I  compute the optimal shape of an annular, thin ribbon as a function of its width. The topological constraint of closing the ribbon determines a relationship between the mean curvature and number of wrinkles that prevents a complete relaxation of the compression strain induced by swelling and buckles the ribbon out of the plane. These results are then applied to thin, buckled disks, where the expansion works surprisingly well. I identify a critical radius above which the disk in-plane strain cannot be relaxed completely. 
\end{abstract}


\maketitle

\section{Introduction}
In nature, elastic instabilities determine the three-dimensional structure of a variety of surfaces, including leaves \cite{nechaev01, sharon07,marder03}, flowers \cite{marder06}, torn plastic sheets \cite{swinney02,audoly03}, growing tissues \cite{drasdo,amar,amar09}, and, possibly, graphene sheets \cite{flagg} . Harnessing the elasticity of thin films artificially, however, remains both a theoretical and experimental challenge \cite{physicstoday, kamien}. In recent work, Klein \textit{et al.} \cite{klein07} have designed thin films that swell isotropically, but inhomogeneously. This creates a residual stress that is relieved by buckling out of the plane, leading to the possibility of controlling three-dimensional structure by patterning a two-dimensional surface. The problem of determining the optimal three-dimensional membrane shape from the underlying swelling pattern, however, remains a poorly understood but critical component of engineering functional thin membrane structures.

For small thickness $t$, membranes exhibit large deformations at relatively small cost compared to planar strain, and one generically expects that they will bend to eliminate most of that strain. Determining the resulting shape has been associated with the mathematical problem of isometric embedding \cite{nechaev01,marder03,marder06} -- how does one determine the buckled shape if one knows only how far apart points on the surface are supposed to be? When are the resulting shapes uniquely determined?
When the buckling induces positive Gaussian curvature, it turns out that this is always possible and the buckled shape is unique \cite{embeddingbook}.
For negatively-curved surfaces, on the other hand, the situation is murkier. Hilbert's theorem and its modern generalizations may forbid the vanishing of the strain \textit{globally} even though the strain can always be accommodated by buckling \textit{locally} \cite{embeddingbook}. It is also known numerically that some isometric embeddings are not unique \cite{audoly03}. However the precise conditions determining whether a strain-free embedding exists and is unique are not at all known in general.

\begin{figure}
\includegraphics[width=0.45 \textwidth]{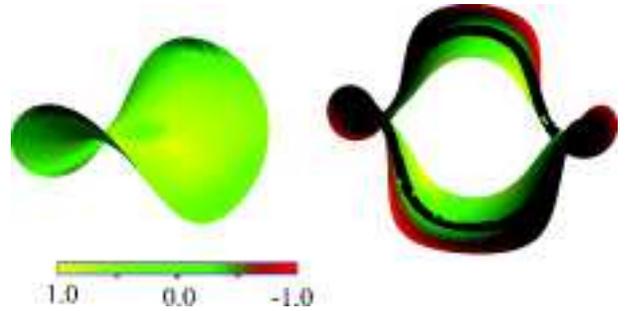}
\caption{A typical disk with prescribed $K = -1/R^2$ (left) and an annular ribbon with three wrinkles (right), color-coded according to Gaussian curvature normalized to lie between $-1$ and $1$. The solid line on the ribbon is a curve of zero mean curvature. The 6-fold symmetry of the Gaussian curvature is apparent at the ribbon edge.}
\label{fig:coords}
\end{figure}

In this letter, I study the physical manifestation of this geometrical frustration for thin, annular ribbons and disks that have been buckled by inhomogeneously swelling (for example, see figure \ref{fig:coords}). 
The goal is to answer questions about the existence and uniqueness of buckled metrics with negative curvature in this restricted set of geometries, and determine how these properties physically manifest themselves in buckled ribbons and disks.
I will focus on the case of excess swelling on the outer edge of the ribbon resulting in a constant, negative Gaussian curvature. I find optimal ribbon shapes as the width increases, and identify a topological constraint that limits the availability of strain-free embeddings. These results are then applied to buckled disks with constant, negative curvature, where I present evidence for a critical radius above which the inhomogeneous strain can no longer be accommodated. Below the critical radius, the disk shape is uniquely determined by the metric.

\section{General considerations}
We consider a model thin film described by coordinates $\textbf{x}_0$ in its flat, unswelled state. We assume either that there is a controlled growth process, such as cell division in a biological tissue \cite{drasdo, amar,amar09}, or the inhomogeneous swelling of a polymer gel \cite{klein07}, after which the film shape is given by $\textbf{X}(\textbf{x}_0)$. The three-dimensional elastic energy can be described in terms of a Cauchy-Green strain tensor, $\Lambda^i_{~j} = \partial \textbf{X}^i/\partial \textbf{x}_0^j = \partial_j \textbf{X}^i$. We assume there is a unique energy minimum with respect to $\Lambda^i_{~j}$. The question is to determine whether there is an $\textbf{X}(\textbf{x}_0)$ for which this $\Lambda^i_j$ exists.
Consider the case of swelling an isotropic polymer gel. For the homogeneous swelling, $\Lambda^i_{~j} = \Omega \delta^i_j$, were $\Omega$ depends on the cross-link density as well as other material parameters \cite{treloar}. When the cross-link density in inhomogeneous, as in Ref. \cite{klein07}, I assume $\Omega$ becomes of a function of the \textit{local} density. For a thin film, dimensional reduction reduces the free energy to a \textit{two-dimensional} membrane energy for the neutral surface of the film \cite{efrati08},
\begin{equation}
E = \int dA~\frac{1}{2} \left[\lambda \left(\textrm{tr} \gamma\right)^2 + 2 \mu \textrm{tr} \left(\gamma^2\right) + \kappa (H - c_0)^2 \right],
\end{equation}
where $H$ is the mean curvature of the film, given by the average of the two principle curvatures, and $\gamma$ is the strain. In this letter, I will neglect a the possibility of spontaneous curvature, which would arise if $\Omega$ had a gradient across the film thickness, by setting $c_0 = 0$.
In general, the strain can be written in the form $\gamma_{i j} = g_{i j} - \bar{g}_{i j}$, where $dl^2 = g_{i j} dx^i dx^j$ is the two-dimensional metric of the surface and $d\bar{l}^2 = \bar{g}_{i j} dx^i dx^j$ is the prescribed two-dimensional metric. The trace is the usual one on curved surfaces, $\textrm{tr} \gamma = \bar{g}^{i j} \gamma_{i j}$, save that we use the prescribed metric for the contraction. Similarly, $\textrm{tr} (\gamma^2) = \bar{g}^{i l} \gamma_{i j} \bar{g}^{j k} \gamma_{k l}$.
For a membrane of thickness $t$, $\kappa/\lambda \propto t^2$. Therefore, the balance between bending and stretching energies is dominated by stretching and a reasonable approximation is to minimize the stretching energy completely, allowing the membrane to bend in whatever way necessary to do that. In general we should expect that the moduli vary with the swelling, but will neglect this complication in the following analysis.

For isotropic swelling, the reference metric is $d\bar{l}^2 = \Omega(r) \left[d\bar{r}^2 + \bar{r}^2 d\bar{\theta}^2\right]$, where $(\bar{r},\bar{\theta})$ describe polar coordinates on the unbuckled membrane. We will find it useful, however, to choose a coordinate system in which the \textit{buckled} membrane has a particularly simple mathematical representation. In Gaussian normal coordinates $(r,\theta)$, the metric will be $dl^2 = dr^2 + \rho^2(r,\theta) d\theta^2$, where $\bar{\rho} = r + \ln \Omega(r)^{-1/2}$. These coordinates generalize the polar coordinates of a flat disk or ribbon: $r$ measures the arc length along radial geodesics emanating from the center of the disk or inner ribbon boundary while $\theta$ is the azimuthal angle around the disk or ribbon. These coordinates always exist in disk or ribbon geometries \cite{embeddingbook}. The boundaries are from $r=r_0-w/2$ to $r=r_0+w/2$ for ribbons and $r=0$ to $r=w$ for a disk of radius $w$.
The Gaussian curvature is given entirely in terms of the metric by $K = -\partial_r^2 \rho/\rho$. For positively curved disks, $\rho$ increases slower than $r$; for negatively curved disks, it grows faster. For $K=-1/R^2$, for example, $\rho = R \sinh(r/R)$ which is linear near the origin and grows exponentially toward the edges.
For the remainder of this letter, I will specialize to prescribed metrics with constant, negative curvature, though my methods will be applicable to a larger class of curvature distributions.

Why can negatively-curved metrics not always be embedded? The classic no-go theorem was given by Hilbert, who showed that there is no surface of constant, negative curvature that is both \textit{smooth} and infinite area \cite{embeddingbook}. For axisymmetric shapes, this is apparently manifested physically by the divergence of the bending energy, which always occurs when $\partial_r \rho \sim 1$ \cite{physicstoday}. This is a compelling physical picture: the membrane is nearly strain-free until the curvature grows too large, after which the bending energy forces the film to have in-plane strain no matter how small the thickness. The situation, however, is more complex than it first appears. A surprising theorem of Nash demonstrated that a surface can be bent to have any metric but that the curvature may not be continuous \cite{nash}. In particular, this implies that the bending energy need not diverge in general \cite{footnote}. These theorems do not, unfortunately, lead to a general criteria one can use to determine when a negatively-curved metric can be strain-free.

The key to our analysis is to examine the curvature more closely. The membrane curvature is described by the tensor $h_{i j} = \hat{n} \cdot \partial_i \partial_j \textbf{r}$, where $\textbf{r}(r,\theta)$ describes the embedding of the surface as a function of the coordinates, $\hat{n}$ is the unit normal to surface and the indices $i=r,\theta$. The mean curvature $H= (1/2) (h_{r r} + h_{\theta \theta}/\rho^2)$ and $(h_{r r} h_{\theta \theta} - h_{r \theta}^2)/\rho^2 = K$ is the Gaussian curvature, $K$. This formula for $K$, when combined with $K = - \partial_r^2 \rho/\rho$, yield a surprising result for isometric embeddings: the entire shape can be determined by integrating the evolution equation \cite{janet, embeddingbook}
\begin{equation}\label{eq:evolution1}
\partial_r^2 \textbf{r} = \hat{n} \frac{\left(\hat{n} \cdot \partial_r \partial_\theta \textbf{r}\right)^2 - \rho \partial_r^2 \rho}{\hat{n} \cdot \partial_\theta^2 \textbf{r}}.
\end{equation}
This equality, also known as Gauss' \textit{theorem egregium}, is a
second order evolution equation requiring initial conditions $\textbf{r}(r_0,\theta)$ and $\partial_r \textbf{r}(r_0,\theta)$, since $\hat{n} = \partial_r \textbf{r} \times \partial_\theta \textbf{r}/\rho$. This initial data, however, is further constrained by compatibility with the metric: (i) $\partial_r \textbf{r}^2 = 1$, (ii) $\partial_\theta \textbf{r}^2 = \rho^2$ and (iii) $\partial_r \textbf{r} \cdot \partial_\theta \textbf{r} = 0$. This is sufficient to completely constrain $\partial_r \textbf{r}$ up to an overall sign and, by equation (\ref{eq:evolution1}), the rest of the surface as well.

\begin{figure}
\includegraphics[width=0.45 \textwidth]{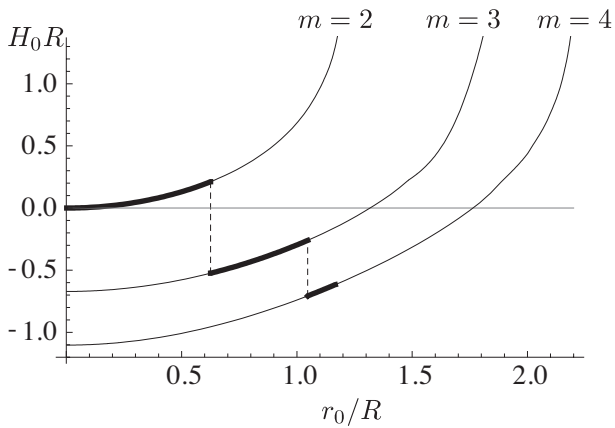}
\caption{$H_0 R$ as a function of $r_0/R$ for different values of $m$. The thick line follows the mean curvature of the optimal ribbon of radius $r_0$ according to equation \ref{eq:optimal} up to $m=4$. The transitions shown are at $r_0/R \approx 0.63$ ($m=2$ to $3$) and $r_0/R \approx 1.05$ ($m=3$ to $4$).}
\label{fig:H0}
\end{figure}

To determine the shape of wrinkled membranes, the evolution equation (\ref{eq:evolution1}) can be recast as an evolution for the $h_{i j}$, yielding
\begin{eqnarray}
\partial_r h_{r \theta} &=& \partial_\theta h_{r r} - \frac{\partial_r \rho}{\rho} h_{r \theta}\nonumber\\
\partial_r h_{\theta \theta} &=& h_{r r} \rho \partial_r \rho + \partial_\theta h_{r \theta} + \frac{\partial_r \rho}{\rho} h_{\theta \theta} - \frac{\partial_\theta \rho}{\rho} h_{r \theta}\label{eq:evolution2}\\
\rho \partial_r^2 \rho &=& h^2_{r \theta} - h_{\theta \theta} h_{r r}.\nonumber
\end{eqnarray}
Equations \ref{eq:evolution2} are a form of the Gauss-Codazzi system \cite{embeddingbook}.

If we use the last equation to solve for $h_{r r}$ we will have to contend with the fact that $h_{\theta \theta}$ may vanish. This is precisely what happens for axisymmetric surfaces, where $h_{\theta \theta} \rightarrow 0$ when $\partial_r \rho \rightarrow 1$ implies a diverging mean curvature at one edge. 
This problem is exacerbated in disks, where $\partial_r \rho \ge 1$ and no axisymmetric embedding exists. As it happens, however, $h_{\theta \theta}$ can vanish without $H$ diverging.
Since $2 H = [h_{r \theta}^2 + (h_{\theta \theta}/\rho)^2-\rho \partial_r^2 \rho]/h_{\theta \theta}$, finite $H$ requires that the components of the curvature tensor lie on a circle. In other words,
\begin{eqnarray}\label{eq:curvs}
h_{r \theta} &=& \sqrt{\rho^2 H^2 + \rho \partial_r^2 \rho} \cos \alpha,\\
h_{\theta \theta}/\rho &=& \rho H - \sqrt{\rho^2 H^2 + \rho \partial_r^2 \rho} \sin \alpha,\nonumber
\end{eqnarray}
for some function $\alpha(r,\theta)$. This, in turn, implies
\begin{equation}\label{eq:hrr}
h_{r r} \rho = \rho H + \sqrt{\rho^2 H^2 + \rho \partial_r^2 \rho} \sin \alpha.
\end{equation}
This is akin to coordinate singularities in general relativity; the problem is not with the surface but with the coordinate system on the surface. Notice the following interesting fact, however: we can choose \textit{any} $h_{r r}$ and integrate equations (\ref{eq:evolution2}) to find the resulting metric without this difficulty. Apparently, the entire surface is determined by our choice of initial conditions and an arbitrary function, $h_{r r}$. The task will be to choose $h_{r r}$ such that the resulting $\rho$ agrees with a prescribed metric $\bar{\rho}$ as closely as possible; as it turns out, this allows us to side step the issue of vanishing $h_{\theta \theta}$.

\begin{figure}
\includegraphics[width=0.45 \textwidth]{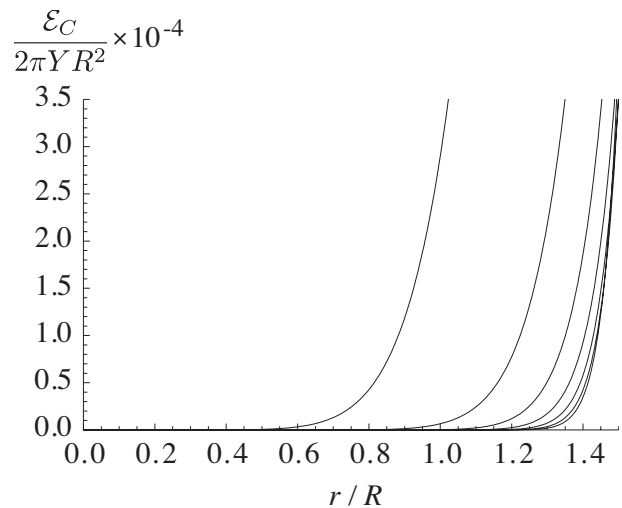}
\caption{Plot of normalized strain energy per unit radial length, $\mathcal{E}_C(r)$ for a negatively curved disk with $h_{r r}$ increasing from zeroth order to eleventh order. The curves converge toward the rightmost curve with increasing order.}
\label{fig:strain}
\end{figure}

\section{Annular ribbons}
I will first consider annular ribbons of width $w$ and thickness $t \ll w$. I will think of these ribbons as having been cut from a disk with metric $\bar{\rho} = R \sinh(r/R)$ from $r=r_0 - w/2$ to $r = r_0 + w/2$. This will greatly simplify our discussion of disks in the next section. Since the ribbons are thin, we solve the curvature evolution equations by expanding $h_{r r} = \sum_n h_{r r,n} (\delta r/R)^n$ and $\rho = \sum_n \rho_n (r/R)^n$. This will yield a unique, perturbative solution in powers of $\delta r = r-r_0$ for the shape.

To make progress, we must make one further simplification when the in-plane strain does not vanish. The strain contribution to the energy is controlled by $\Delta =(\rho^2 - \bar{\rho}^2)/\bar{\rho}^2$; after all, if $\Delta=0$ then the actual metric equals the prescribed metric and $\gamma_{i j}=0$. Calculating the strain when $\Delta \ne 0$ can be quite difficult, however. If we assume $\Delta$ is small, we can expand all the strain components in powers of $\Delta$ and the dominant term in the strain energy will scale with $\Delta^2$. Clearly, minimizing $\Delta^2$ is sufficient to find a strain-free configuration when one exists; I also expect that it provides an adequate approximation to the shape even when one does not exist. To obtain an explicit expression for the strain, I follow Audoly and Boudaoud \cite{audoly03} by setting all but one strain component to zero. I choose $\gamma_{r r} = \gamma_{r \theta} = 0$, leaving $\gamma_{\theta \theta} = \Delta$. This approximation limits us, however, to regimes in which either in-plane strain or bending energy dominates the energetics since it probably incorrectly captures how these two contributions to the energy would balance. In the case of buckled disks, we will only be concerned with the existence and shape of surfaces with $\Delta=0$ so this should not be a problem.

As long as $w \gg t$ the strain energy ensures that $\rho_0 = \bar{\rho}_0$ and $\rho_1 = \bar{\rho}_1$ -- that is the metric agrees with the prescribed metric up to first order in $w/R$. The lowest order terms in the energy are then given by
\begin{eqnarray}
\frac{E}{Y R^2} &=& \frac{\pi}{40} \left[\int \frac{d\theta}{2 \pi} \frac{R^4}{\bar{\rho}_0^2} \left(\rho_2 - \bar{\rho}_2\right)^2\right] \left(\frac{w}{R}\right)^5\\
& & + \pi \left(\frac{t}{R}\right)^2 \left(\frac{w}{R} \right) \frac{\bar{\rho}_0}{R} \left[ \int \frac{d\theta}{2 \pi} H^2(r_0,\theta) R^2 \right],\nonumber
\end{eqnarray}
where $Y$ is the Young's modulus. Due to our approximation for the strain, one should not play too close attention to the numerical factors -- I've included them here for completeness. The scaling of the compression and bending energies with respect to $w/R$ allow us to identify two regimes: $w/R \ll (t/R)^{1/2}$, for which the bending energy term dominates, and $w/R \gg (t/R)^{1/2}$, for which the compression term dominates. The former regime was studied in reference \cite{marder06}. Since the annular ribbons shapes are determined by minimizing their bending energy, they are flat ($H(r_0,\theta)=0$) but circular strips.

When $w/R \gg (t/R)^{1/2}$, the dominance of the compression strain results in $\rho_2 = \bar{\rho}_2$. After setting $\rho_2$, the next term in the compression energy scales as $(w/R)^7$. More generally, if we choose $\bar{\rho}_n = \rho_n$, the dominant term in the compression energy scales as $(w/R)^{3+2 n}$. From this, we identify a cascade of regimes $(t/R) \ll (t/R)^{1/2} \ll (t/R)^{1/4} \ll \cdots$ that set the relative balance of compression energy and bending energy order by order. When $w/R \gg (t/R)^{1/4}$, for example, the compression energy requires us to choose $\rho_3 = \bar{\rho}_3$. We then obtain an approximation to the ribbon shape  by minimizing the compression strain up to some power of $w/R$, then set the remaining terms by minimizing the bending energy. This basic picture will not be changed by a more careful evaluation of the strain energy, and if the ribbons were not closed annuli the problem of optimal shape would be solved. In the limit of $t \rightarrow 0$, in particular, we conclude that we should systematically set the coefficients $\rho_n = \bar{\rho}_n$ in the ground state. This analysis implies that we are able to choose any $H$ and $\alpha$ arbitrarily for our initial conditions, from which the entire expansion in powers of $r$ is determined. This analysis conforms with the notion that there is no local obstruction to finding strain-free configurations.

For annular ribbons, however, there is an obstruction to our ability to minimize the compression energy term to third order and above: the ribbon must be periodic and not twisted. This constraint is encoded by the winding numbers of the three Euler angles of the coordinate frame defined by $\partial_r \textbf{r}$, $\partial_\theta \textbf{r}$ and $\hat{n}$. The unit tangent vector of a curve of constant $r$ can be written as $\hat{t} = \cos \theta \sin \phi \hat{x} - \cos \theta \cos \phi \hat{y} - \sin \theta \hat{z}$. The remaining Euler angle, $\psi$ defines angle made by the surface normal $\hat{n}$ around the axis defined by $\hat{t}$: $\hat{n} = (- \sin \psi \cos \phi + \sin \theta \cos \psi \sin \phi) \hat{x} + (- \sin \psi \sin \phi - \sin \theta \cos \psi \cos \theta) \hat{y} + \cos \theta \cos \psi \hat{z}$. Using these definitions, we have $\sin \psi \theta' - \cos \theta \cos \psi \phi' = - \rho_1$ from the metric, where the prime indicates differentiation with respect to the azimuthal coordinate. These angles are related to the curvatures by
\begin{eqnarray}
\psi' &=& \tan \theta \left[ \rho_1 \cos \psi - \sin \psi h_{\theta \theta}/\rho_0 \right] - h_{r \theta}\nonumber\\
\theta' &=& - \cos \psi h_{\theta \theta}/\rho_0 - \sin \psi \rho_1\\
\phi' &=& \frac{\cos \psi \rho_1 - \sin \psi~h_{\theta \theta}/\rho_0}{\cos \theta}.\nonumber
\end{eqnarray}
The constraint states simply that $\Delta \phi = \phi(2 \pi) - \phi(0) = 2 \pi$, $\Delta \psi=0$, and $\Delta \theta = 0$, or some other combination where one Euler angle gives $2 \pi$ and the others give zero.

We look for a solution of these equations of the form $H = H_0 \sin (m \theta)$, $\alpha = m \theta$ and $\rho_2 = \bar{\rho}_2$ in equation (\ref{eq:curvs}), choosing the number of wrinkles $m$, and $H_0$, to satisfy the winding number constraints. The numerically determined relationship is shown in figure \ref{fig:H0}. Apparently, $H_0$ is determined completely by $m$ and $\rho_1$, and it is no longer possible to set the curvature to zero. Moreover, this relationship must be true not only for ribbons but for all curves of constant $r$ on a membrane. Notice that $H_0 = 0$ along $r=r_0$ whenever $m=\bar{\rho}_1 + 1$. Figure \ref{fig:coords} displays a ribbon with $m=3$ and $H_0 = 0$ at $r=r_0$.

With these results in hand, we consider again the case that $w/R \gg (t/R)^{1/4}$ in which we are required to choose a minimal value of $\rho_3$. The evolution equation for $\rho$ determines $\rho_3$ in terms of the curvatures at some $r=r_0$. Assuming $H_0 R$ is small, we expand the compression energy to quadratic order in the mean curvature, yielding
\begin{eqnarray}\label{eq:optimal}
& &\frac{E_C}{2 \pi Y R^2} = \frac{1}{576} \left(\frac{R}{\bar{\rho}_0}\right)^3 \left(\frac{w}{R}\right)^7\\
& &\times \left[ \left(2 \bar{\rho}_2 R - \sqrt{2 \bar{\rho}_0 \bar{\rho}_2} H_0 R\right)m - 3 R \left(\bar{\rho}_1 \bar{\rho}_2+\bar{\rho}_0 \bar{\rho}_3\right)\right]^2.\nonumber
\end{eqnarray}
This expression can be minimized subject to the topological constraint. The results of this minimization are displayed as a thick solid line in figure \ref{fig:H0} in terms of the mean curvature. In particular, I find that $m=2$ is preferred for small radii, $m=3$ is energetically preferred when $r \approx 0.63 R$, and $m=4$ when $r \approx 1.05 R$. This trajectory differs from the optimal ribbon in the regime $(t/R)^{1/2} \ll w/R \ll (t/R)^{1/4}$, when bending energy dominates over compression strain. In that case, the number of wrinkles follows the trajectory in figure \ref{fig:H0} that minimizes $H_0^2$. This fundamental incompatibility arises from the topologically determined relationship between $H_0$ and the metric -- if we were to cut the ribbon to relax this constraint, a ground state with $H_0 = 0$ and $\rho_3 = \bar{\rho}_3$ would yield a non-integer $m$ according to equation \ref{eq:optimal}.

\begin{figure}
\includegraphics[width=0.45 \textwidth]{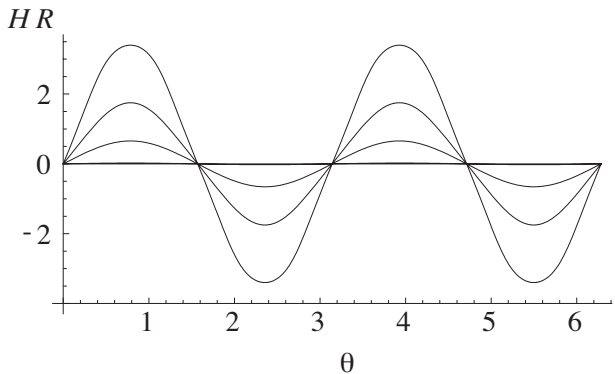}
\caption{$H R$ for the disk of figure \ref{fig:coords} for $r=R,3 R/2$, and $2 R$.}
\label{fig:Hcurvatures}
\end{figure}

\section{Buckled Disks}
We now turn our attention to disks. Since every curve of constant $r$ on a disk satisfies the same topological constraint that ribbons do, we can recover the limit of a disk by taking $r_0 \rightarrow 0$ in figure \ref{fig:H0}. This implies that $\bar{\rho}_1 \rightarrow 1$. Notice that when $m=2$, $H(0) = 0$ and the disk is smooth at the center. On the other hand, if $m>2$ on the disk as $r_0 \rightarrow 0$, the surface will not be smooth at $r=0$ since $H = H_0(0) \sin(m \theta)$ -- the result is a singularity at the disk center akin to that of a cone.

We again apply an expansion in powers of $r/R$. We will assume that the $t$ is small enough that the compression strain dominates to all orders of $r/R$. This also ensures that the resulting shape will be independent of the approximation used to evaluate the strain since we need only minimize $\Delta^2$. Immediately, we see that $\rho = \sum_n \rho_n (\delta r/R)^n$ agrees with $\bar{\rho} = \sum_n \bar{\rho}_n (\delta r/R)^n$ up to second order when $h_{r r}$ is given by equation (\ref{eq:hrr}) evaluated on $r=r_0$. Using this expansion, I find that $\alpha = 2 \theta$ as $r_0 \rightarrow 0$ as expected from our analysis of figure \ref{fig:H0}. Negatively-curved disks are saddle-shaped near their centers regardless of the magnitude of Gaussian curvature.  This result is anticipated by a stability analysis of growing tissues \cite{amar} and the fact that saddles are canonical examples of negatively-curved surfaces. Continuing in this manner, 
\begin{eqnarray}
h_{r r} &=& \left[1+(1/2) \delta r^2/R^2 + (1/6) \delta r^4/R^4\right] R^{-1} \sin (2 \theta)\nonumber\\
& & - (1/24) \delta r^4/R^4 \sin(6 \theta)\label{eq:hrrdisk}
\end{eqnarray}
up to fourth order, implying that $\rho$ agrees with the prescribed metric up to seventh order. The resulting shape is shown in figure \ref{fig:coords}.
It is worth noting that for small but finite $t/R$, we might expect deviations from this shape due to small, but nonzero, contributions from the bending energy. In this case, the resulting deviations of the Gaussian curvature are fourfold symmetric. In fact, $K \propto \sin^2(2 \theta)$, which is a consequence of its quadratic dependence on the $h_{i j}$ and the membrane's saddle shape. This four-fold symmetric deviation is apparent in experiments on non-Euclidean disks \cite{klein07} and is universal for negatively-curved disks with axisymmetric $\bar{\rho}$.

This perturbative construction can be continued to higher orders, forcing the introduction of higher oscillations in $\theta$. The data in figure \ref{fig:strain} is shown up to eleventh order in $r/R$, after which the deviation of the metric appears to converge to a universal curve. I have confirmed this convergence up to fifteenth order in $r/R$. It is interesting that the perturbative expansion is quite good, as measured by $\Delta^2$, even with relatively few terms, as long as $r < R$.
From figure \ref{fig:strain}, above a radius, $r \approx 1.2 R$, $\Delta^2$ does not decrease with increasing order of $r/R$ though it does below this radius.

It is enlightening to consider these results in light of several rigorous mathematical results on negatively curved surfaces. Since a surface can be deformed to agree with any prescribed metric in a finite region \cite{janet} we conclude that the unique expansion in equation (\ref{eq:hrrdisk}) must have a finite (nonzero) radius of convergence around the origin. Hilbert's theorem, however, ensures that this expansion fails to converge at large enough $r$ even at $t \rightarrow 0$ since, if it did, the surface would have infinite area and constant negative curvature.
A more recent mathematical result does provide an existence proof (but not a construction) that a negatively-curved disk can be embedded up to a critical radius $r_c$ \cite{embeddingbook, hong}. According to this result, above this radius $|K|$ must either decrease faster than $1/r^2$ or develop zeros no matter how $h_{r r}$ is adjusted. From figure \ref{fig:strain}, I find $r_c \approx 1.2 R$ as the critical radius for constant curvature.
These theorems, taken together, imply the disk becomes non-analytic at $r=r_c$ as $t \rightarrow 0$. This follows immediately from the fact that $\rho^2-\bar{\rho}^2 = 0$ for $r < r_c$ and is nonzero for $r > r_c$.

Notice that figure \ref{fig:Hcurvatures} indicates that $H = H_0(r) \sin (2 \theta)$ on a disk to excellent approximation. We glean further insights by considering the topological constraint on curves of constant $r$ on the disk. As it turns out, $\int d\theta  H_0(r,\theta)^2/(2 \pi)$ on the disk agrees quite well with the value predicted by the $m=2$ branch of figure \ref{fig:H0}.
This prediction works until the $m=2$ branch ends at $r \approx 1.2 R$, beyond which the constraint cannot be satisfied by solutions of the form $H=H_0 \sin(2 \theta)$. This gives an independent corroboration of the value of critical radius found by perturbative expansion in powers of $r/R$. 
Unlike the case of axisymmetric surfaces, the mean curvature does not diverge at this critical radius. This roughly matches our intuition from Nash's theorem that the curvature need only be discontinuous at worst, not divergent. It is interesting to speculate that the curvature would become discontinuous precisely at $r=r_c$ as $t \rightarrow 0$ if we pursued an isometric embedding beyond the critical radius.

\begin{figure}
\includegraphics[width=0.45 \textwidth]{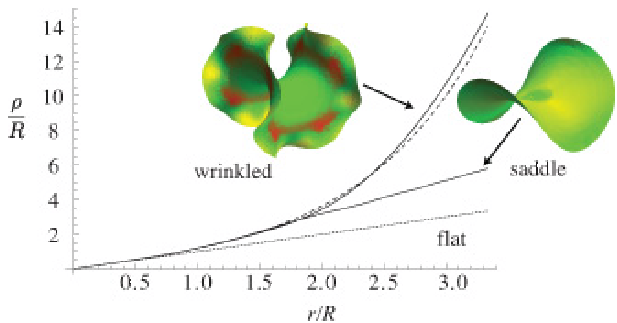}
\caption{$\rho(r)$ for the saddle (up to order $r^4$) with and without wrinkled edges, compared to the prescribed metric $\bar{\rho}(r) = R \sinh(r/R)$ (dashed). For comparison, $\rho(r) = r$ (dotted), a flat disk, is shown.
}
\label{fig:rho}
\end{figure}

\section{Conclusion}
What happens above the critical radius? 
A hint comes from the fact that better agreement between disk shape and prescribed metric can be made if we allow additional subwrinkles. 
In figure \ref{fig:rho}, I compare the constant curvature saddle to a disk on which additional subwrinkles begin abruptly at $r/R = 4/3$ with $m = 10$. This was chosen to illustrate the significant benefit in strain achieved by adding subwrinkles.
Subwrinkling is often observed in negatively-curved leaves \cite{sharon07, marder03}, strips \cite{swinney02}, flowers \cite{marder06}, and disks \cite{klein07}. It is tempting to speculate that the zero compression strain trajectory of figure \ref{fig:H0} also indicates the preferred number of wrinkles as a function of radius on a disk. Unfortunately, this does not account for the energy cost of changing the number of wrinkles and so it is unlikely that disks will exhibit all the $m$ as they do for ribbons. Further insights will be needed.

In summary, I have introduced new methods to compute optimal shapes of annular ribbons and thin disks. This theory identifies a cascade of shape regimes as a function of ribbon width and makes definite predictions for the resulting optimal shapes. In particular, I consider the new regimes $(t/R)^{1/2} \ll w/R \ll (t/R)^{1/4}$ and $(t/R)^{1/4} \ll (w/R) \ll (t/R)^{1/6}$. The novel feature in these regimes are the interplay between topology and curvature, which constrains the number of wrinkles and mean curvature. These topological constraints also apply to closed curves on thin disks, and can be used as good approximations to the shape of negatively curved disks.

One obvious limitation of the theory is the computation of the strain energy, which sets two strain components to zero. This is most assuredly incorrect. In the present work, I have limited consideration to regimes in which there is a clear dominance of either the compression or the bending energy order by order in perturbation theory. Therefore, minimization of the compression strain amounts to setting the appropriate power of the metric equal to its prescribed value. This assumption breaks down beyond the critical radius $r_c$ on a disk, as well as in the cross-over regions between two regimes of an annular ribbon -- for example, when $w/R \sim (t/R)^{1/4}$ -- where compression and bending energies are roughly balanced. The ribbon shape in these cross-over regions remains elusive.

Though this letter is limited to disks and ribbons with \textit{constant} negative curvature, the methods of analysis should be directly applicable to more general metrics. Future work will focus on the defining the precise conditions under which strain-free shapes exist, understanding which metrics uniquely specify a shape, and making a more direct connection between experimentally-controlled parameters such as the cross-link density and the buckled film shape. Finally, the predictions made by this theory can be tested on thin films with prescribed inhomogeneous swelling patterns.

\acknowledgements
I am grateful to G. Grason and R.D. Kamien for useful discussions and a reading of the manuscript. I thank the Aspen Center for Physics, where a portion of this work was done.


\begin{thebibliography}{0}
\bibitem{footnote}
The precise statement of Nash's theorem is that there is exists an isometric, $C^1$ immersion of any surface in $R^3$. In the case of a singularity of the curvature, the tangent vectors are discontinuous and the surface would not even be $C^1$ at the singularity.
\end{thebibliography}
\end{document}